\documentclass[preprint2,times,tighten]{aastex6}

\pdfoutput=1 
\usepackage{amsmath,amstext}
\usepackage[all]{hypcap} 
\usepackage{verbatim}

\shortauthors{Oklop\v{c}i\'c, Hirata \& Heng}

\begin{document}

\title{How does the shape of the stellar spectrum affect the Raman scattering features in the albedo of exoplanets?}

\author{Antonija Oklop\v ci\'c\altaffilmark{1,2}, Christopher M. Hirata\altaffilmark{3} and Kevin Heng\altaffilmark{4}}
\email{antonija.oklopcic@cfa.harvard.edu}
\altaffiltext{1}{California Institute of Technology, 1200 East California Boulevard, MC 249-17, Pasadena, California 91125, USA}
\altaffiltext{2}{Institute for Theory and Computation, Harvard-Smithsonian Center for Astrophysics, 60 Garden Street, MS-51, Cambridge, Massachusetts 02138, USA}
\altaffiltext{3}{Center for Cosmology and Astroparticle Physics, Ohio State University, 191 West Woodruff Avenue, Columbus, Ohio 43210, USA}
\altaffiltext{4}{Center for Space and Habitability, University of Bern, Sidlerstrasse 5, CH-3012, Bern, Switzerland}

\begin{abstract}
Diagnostic potential of the spectral signatures of Raman scattering, imprinted in planetary albedo spectra at short optical wavelengths, has been demonstrated in research on Solar System planets, and has recently been proposed as a probe of exoplanet atmospheres, complementary to albedo studies at longer wavelengths. Spectral features caused by Raman scattering offer insight into the properties of planetary atmospheres, such as the atmospheric depth, composition, and temperature, as well as the possibility of detecting and spectroscopically identifying spectrally inactive species, such as H$_2$ and N$_2$, in the visible wavelength range. Raman albedo features, however, depend on both the properties of the atmosphere and the shape of the incident stellar spectrum. Identical planetary atmospheres can produce very different albedo spectra depending on the spectral properties of the host star. Here we present a set of geometric albedo spectra calculated for atmospheres with H$_2$/He, N$_2$, and CO$_2$ composition, irradiated by different stellar types ranging from late A to late K stars. Prominent albedo features caused by Raman scattering appear at different wavelengths for different types of host stars. We investigate how absorption due to alkali elements, sodium and potassium, may affect the intensity of Raman features, and discuss the preferred observing strategies for detecting Raman features in future observations.
\end{abstract}

\keywords{molecular processes -- planets and satellites: atmospheres -- radiative transfer -- scattering -- techniques: spectroscopic}
\maketitle
\section{Introduction}

An important observable quantity in exoplanet observations is the flux received from the planet ($F_p$), measured relative to the flux of its host star ($F_*$). This ratio depends on the physical and orbital parameters of the planet-star system, such as the radius of the planet ($R_p$), its distance from the star ($a$), and the orbital phase angle ($\alpha$) with respect to the observer, in the following way:
\begin{equation}
\frac{F_p}{F_*} = A_g \left(\frac{R_p}{a}\right)^2\phi(\alpha) \ \mbox{,}
\label{eq:flux_ratio}
\end{equation}
where $\phi$ denotes the phase function of the planet.
Most of the information about atmospheric properties is contained in the wavelength-dependent quantity $A_g$, called the geometric albedo. At short optical wavelengths, thermal radiation of planets is considered to be negligible compared to reflected starlight. Therefore, by measuring the optical albedo of a planet, we are effectively measuring the reflective properties of its atmosphere and/or surface \citep[e.g.][]{Marley1999,Hu2012}. 

Measurements of the geometric albedo in the optical wavelength range have been obtained for a few dozen exoplanets to date \citep[e.g.][]{Rowe2008, Snellen2009, Christiansen2010, Desert2011, Kipping2011, CowanAgol2011, Demory2013}, mostly via photometric secondary eclipse observations with telescopes like \textit{MOST}, \textit{CoRoT}, and \textit{Kepler}. Even though these broadband observations provide just one number for each planet---the value (or an upper limit) of the albedo averaged over the entire bandpass of the telescope---the albedos obtained so far span a broad range of values, pointing to a large diversity in the properties of exoplanet atmospheres. Most hot Jupiters have low values ($\lesssim0.1$) of $A_g$ \citep{HengDemory2013}; however, there are some notable exceptions, such as \textit{Kepler}-7b \citep{Latham2010} with $A_g = 0.35$ \citep{Demory2011}. \citet{Martins2015} observed the reflected light of a non-transiting hot Jupiter 51~Peg~b in the wavelength range $\sim 380 - 690$~nm using ground-based high-resolution spectroscopy and found tentative evidence for a high average value of the geometric albedo\footnote{There is an inherent degeneracy between the radius of the planet and its albedo (see \autoref{eq:flux_ratio}). The results of \citet{Martins2015} suggest geometric albedo values greater than unity for a radius of 1.2 R$_{\mathrm{Jup}}$. The authors therefore conclude that 51~Peg~b may be an inflated hot Jupiter with a radius of 1.9~R$_{\mathrm{Jup}}$, assuming $A_g=0.5$. \citet{Birkby2017} present a more detailed discussion of that result, related to their recent mass measurement of 51~Peg~b.}. Super-Earths observed with \textit{Kepler} have been reported to have statistically larger geometric albedos compared to hot Jupiters \citep{Demory2014}. 

\begin{figure}
\centering
\includegraphics[width=0.5\textwidth]{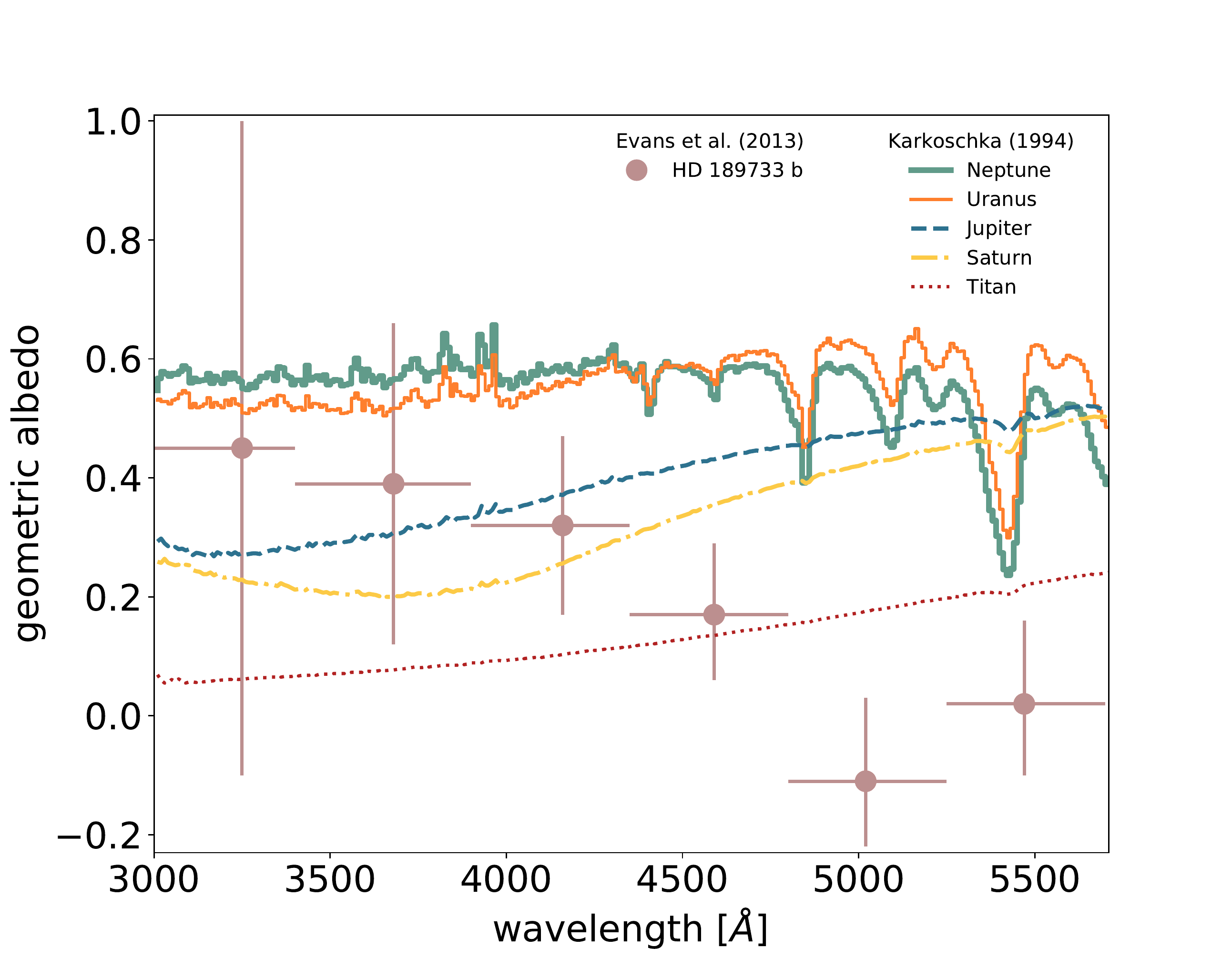}
\caption{Lines show the measured full-disk albedo spectra of the jovian planets and Titan from \citet{Karkoscka1994}. The accuracy of the absolute albedo measurements is $\sim$4\%. The albedo spectra of Uranus and Neptune show prominent spectral features due to atmospheric absorption, as well as the Raman scattering features (peaks resembling emission lines). The spectra of Titan, Jupiter, and Saturn are aerosol-dominated. The measured geometric albedo spectrum of hot Jupiter HD~189733~b from \citet{Evans2013} is shown for comparison.}
\label{fig:obs_albedo}
\end{figure}

\citet{Evans2013} measured the albedo of hot Jupiter HD~189733~b over the wavelength range 290-570~nm using the \textit{STIS} instrument on the \textit{Hubble Space Telescope}. They report the geometric albedo binned in six wavelength channels, thus providing a low-resolution albedo spectrum of this exoplanet (shown in \autoref{fig:obs_albedo}). The measured albedo spectrum of HD~189733~b is broadly consistent with an atmosphere dominated by Rayleigh scattering and sodium absorption, as well as with models containing cloud particles of various compositions and sizes \citep{HengDemory2013, Barstow2014, Heng2014}. The quality of the current data is insufficient to place tighter constraints on the sources of atmospheric opacity.

Nevertheless, albedo spectra of moderate spectral resolution ($R=10^2-10^3$) at short optical wavelengths (3000~\AA~$\lesssim \lambda\lesssim$~6000~\AA) have the potential to provide valuable information about atmospheric properties, complementary to the information available at longer wavelengths \citep{Burrows2008}, which contain strong molecular absorption bands \citep[e.g.][]{Sudarsky2003, Cahoy2010, Morley2015}. The opacity at short optical wavelengths is dominated by Rayleigh scattering on molecules or small particles. Rayleigh scattering produces high values of albedo (a semi-infinite atmosphere with conservative Rayleigh scattering would have $A_g=0.7975$; \citealt{Prather1974}) and a smooth and featureless albedo spectrum \citep{Marley1999, Sromovsky2005b}. However, Rayleigh scattering on molecules is always accompanied by Raman scattering, which does imprint spectral features over the smooth Rayleigh spectrum.

Raman features can be seen in the albedo spectra of many Solar System atmospheres, including the Earth's atmosphere. \autoref{fig:obs_albedo} shows the albedo spectra\footnote{All bodies except Jupiter are observed at small phase angles ($< 3^\circ$ for Saturn and Titan; $<1^\circ$ for Uranus and Neptune), and hence the full-disk albedos are almost equal to the geometric albedos. The measured albedo of Jupiter is about 5\% lower than its geometric albedo because the planet is observed at a phase angle of $\sim10^\circ$.} of giant planets and Titan obtained through ground-based observations by \citet{Karkoscka1994}, who identified between forty and sixty individual Raman features in each spectrum. The strongest features are visible in the spectra of clear, Rayleigh-dominated atmospheres of Uranus and Neptune, in which $\sim 17$\% of incident photons at 4000~\AA\ get Raman scattered, whereas in aerosol-dominated atmospheres of Jupiter and Saturn this fraction is $\sim 5$\% \citep{Karkoschka1998}.

Building upon previous research on the Raman effect in the atmospheres of giant planets in the Solar System \citep[e.g.][]{Sromovsky2005, BetremieuxYelle1999, Courtin1999, Yelle1987, Wallace1972, Belton1971, Belton1973, Price1977, CochranTrafton1978, Cochran1981a, Cochran1981b}, we recently demonstrated how spectral signatures of Raman scattering, imprinted in the albedo spectra of gaseous exoplanets at short optical wavelengths, could be used to obtain information about atmospheric properties: composition, temperature, and the presence of clouds \citep[hereafter Paper~I]{Oklopcic2016}.\defcitealias{Oklopcic2016}{Paper~I} Based on a simple feasibility analysis, we concluded that Raman features could be detected in nearby exoplanets using the next generation of ground-based and spaceborne telescopes.

The intensity of Raman features depends on the properties of both the planetary atmosphere and the incident stellar spectrum. Otherwise identical planetary atmospheres can produce a diverse range of albedo spectra depending on the spectral type of the host star, as we demonstrate in this paper. In \citetalias{Oklopcic2016}, we analyzed the effects of Raman scattering on the geometric albedo for different types of atmospheric models, all irradiated by the solar spectrum. In this paper, we explore how Raman features in the albedo spectrum change with changing the spectral type of the host star, from late A to late K spectral types. By identifying the stars that can host planets with the most prominent Raman features over a given wavelength range, our goal is to help guide the selection of optimal targets for future observations of Raman scattering signatures in exoplanet atmospheres. 

This paper is structured as follows. In \autoref{sec:raman_scattering} we briefly introduce the physics of Raman scattering and its effects on the planetary geometric albedo spectra. In \autoref{sec:methods} we describe the input stellar spectra and the model atmospheres used in our radiative transfer calculations. In \autoref{sec:results} we present the results in terms of the geometric albedo spectra of atmospheres of three different compositions (H$_2$/He, pure N$_2$, and pure CO$_2$) irradiated by five different stellar spectra. We investigate how the presence of atmospheric absorbers---alkali metals sodium and potassium---affects the albedo spectrum and the intensity of Raman features. Furthermore, we discuss the implications of our results  in the context of choosing the optimal observing strategy and wavelength range for a given spectral type of planet-hosting stars. Finally, we summarize our conclusions in \autoref{sec:conclusions}.

\section{Raman Scattering}
\label{sec:raman_scattering}

Raman scattering is an inelastic process in which the energy of the scattered photon changes due to its interaction with the molecular scatterer. Raman scattering is related to Rayleigh scattering, but characterized by smaller cross sections (typically a few percent of the Rayleigh scattering cross section for the same molecule). Even though a minor fraction of the reflected light coming from a planet is due to Raman scattering, this component produces specific spectral features that carry information about atmospheric properties that the dominant (Rayleigh) component does not contain. The intensity of the Rayleigh component is directly proportional to the intensity of the incident light at the same wavelength, and therefore a pure Rayleigh albedo spectrum does not possess any sharp spectral features.

\begin{figure}
\centering
\includegraphics[width=0.31\textwidth]{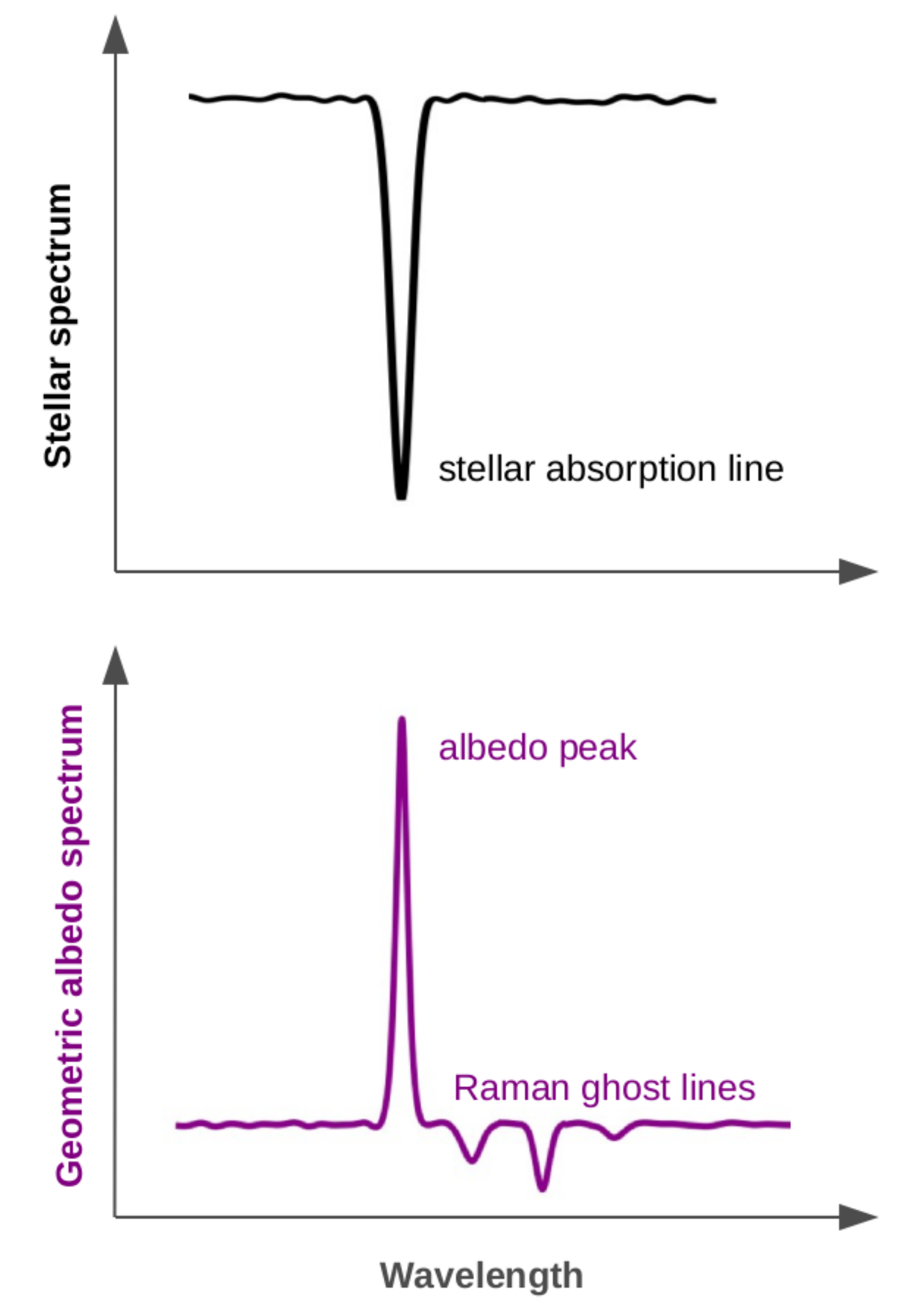}
\caption{Two main types of Raman spectral features---the albedo peaks and the Raman ghost lines---produced by absorption lines in the incident stellar spectrum. Raman albedo peaks appear at the same wavelengths as the stellar absorption lines. Raman ghosts have a well-defined frequency offset with respect to stellar lines that produce them.}
\label{fig:raman_schematic}
\end{figure}

Due to inelastic Raman scattering, the rotational and/or vibrational state of the molecule changes and the energy difference is transferred to the scattered photon, which experiences a change in frequency. All spectral features present in the incident stellar spectrum, such as absorption or emission lines, get shifted in frequency in the Raman-scattered component of reflected light. These frequency shifts result in two main types of Raman features\footnote{In addition to the Raman features imprinted in the radiation intensity or flux, Raman scattering also introduces spectral features in the reflected light polarization spectra. Because Raman scattering is less polarizing than Rayleigh scattering, the lines that get filled-in by Raman-scattered light have a smaller degree of linear polarization than the continuum \citep{Humphreys1984,Aben2001, Stam2002}. Therefore, Raman scattering can affect the polarization spectrum of a planet, which has been proposed as a diagnostic tool for exoplanet characterization \citep[e.g.][]{Stam2004, Stam2008, BuenzliSchmid2009, MarleySengupta2011}.} present in the albedo spectrum of a planet: (1) partial filling-in of absorption lines, resulting in sharp albedo peaks at the wavelengths corresponding to absorption lines in the stellar spectrum, and (2) a series of frequency-shifted absorption lines, resulting in small albedo decrements, called the Raman ghost lines (see \autoref{fig:raman_schematic}; for a more detailed description of how Raman features arise, see \citetalias{Oklopcic2016} and references therein).

Albedo peaks observed at high enough spectral resolution ($R\gtrsim 10^3$) can be quite prominent, especially at wavelengths of strong stellar absorption lines, such as the Ca~II H and K lines in the solar spectrum. Raman ghosts are much fainter, producing changes in the albedo on the order of a few percent. The strength of the albedo peaks depends on the radiation-penetration depth of the atmosphere; thus, by measuring their intensity we can obtain information about the presence and altitude of opaque clouds in the atmosphere. Raman ghost lines can be used to spectroscopically identify the molecule responsible for scattering and thereby infer the bulk atmospheric composition, even in the case of spectrally inactive, symmetric molecules which to not possess a permanent electric dipole moment, like H$_2$ and N$_2$. By comparing the intensity of individual ghost lines, information about the relative population of different molecular levels can be obtained and used to infer the temperature of the atmosphere.

\section{Methods}
\label{sec:methods}

\subsection{Model Atmospheres and Radiative Transfer Calculations}
\label{sec:model_atmospheres}

In order to calculate the geometric albedo spectra, we perform radiative transfer calculations using models of planetary atmospheres. We construct three model atmospheres composed of: 
\begin{enumerate}
\item hydrogen ($^1$H$_2$, 90\% by number) and helium ($^4$He, 10\% by number), with the mean molecular weight of $\mu=2.3$~m$_H$, where m$_H$ is the mass of the hydrogen atom,  
\item nitrogen, $^{14}$N$_2$, with $\mu = 28.0$~m$_H$, 
\item carbon dioxide, $^{12}$C $^{16}$O$_2$, with $\mu = 44.0$~m$_H$.
\end{enumerate}
The surface gravity of the planet in all three cases is set to 1000 cm~s$^{-2}$. 

We assume the temperature-pressure profile that closely resembles the results of \citet{Cahoy2010} obtained for a warm Neptune-like planet at 0.8 au from the host (Sun-like) star, with the atmospheric temperature in the range $\sim 300-1000$~K. We do not compute the temperature profile for each atmosphere in a self-consistent way because the results of \citetalias{Oklopcic2016} show that the atmospheric temperature has only a small effect on the intensity of the most prominent Raman features---the albedo peaks---which are the main focus of this paper. Atmospheric temperature has a much stronger effect on the structure of the less prominent Raman features, the Raman ghosts; however, we do not study them in detail in this paper.

We analyze the propagation of radiation at short optical wavelengths ($\sim 3000-5500$~\AA), for which scattering is expected to be the dominant source of opacity, and ignore absorption (except when otherwise noted) as well as thermal emission of the planet. The wavelength-dependent opacity and single scattering albedo are calculated using cross sections for Rayleigh and Raman scattering of all the relevant species. The calculations used to obtain cross sections for Rayleigh and Raman scattering on hydrogen, helium, and nitrogen are described in \citetalias{Oklopcic2016}. Additionally, in this paper we include the analysis of Raman scattering in CO$_2$-dominated atmospheres, which has not been studied before in the context of exoplanets. The computation of the relevant cross sections for CO$_2$ is described in \autoref{app:co2}. In \autoref{sec:results}, we discuss how the presence of alkali metals in the atmosphere may affect the detectability of Raman features. For the albedo spectra presented there, we included absorption by sodium and potassium as an additional source of opacity. The absorption cross sections used for that analysis are described in \autoref{app:alkali}.

The plane-parallel atmospheres are divided into 49 homogeneous layers, evenly distributed in log-pressure from $10^{-4}$ to 100~bar. The atmospheric conditions (pressure, temperature, opacity, single-scattering albedo) are uniform within each layer and change discontinuously at the layer boundaries called levels. The bottom level is a surface with albedo set to zero. 

The atmosphere is illuminated at the top boundary by a beam of radiation incident at a specified angle. We perform radiative transfer calculations using a discrete ordinate algorithm \textsc{disort} \citep{Stamnes1988}, which we modified to include Raman scattering (see \citetalias{Oklopcic2016} for details). The code calculates the outgoing flux at the top boundary for every wavelength and the specified outgoing angle, which varies at different points on the planetary disk. We find the intensity for nine locations on the disk and use them to compute the total emergent intensity at zero phase angle using the Gaussian quadrature integration method of order $n=6$ \citep{Horak1950, HorakLittle1965,MadhusudhanBurrows2012}. The geometric albedo is then calculated at each wavelength as the ratio of the total emergent intensity of the planet and the incident stellar flux.
 
The radiative transfer code \textsc{disort} uses a scalar representation of light instead of a full (Stokes) vector representation. This approximation is often used in order to reduce the complexity and the computational cost of radiative transfer calculations. Neglecting the effects of light polarization in the treatment of atmospheric scattering can cause errors in the calculated flux, and consequently in the geometric albedo of the planet, on the order of a few percent \citep[e.g.][]{Chandrasekhar1950, Mishchenko1994, Lacis1998}. The differences in flux calculated with and without taking polarization into account are largest for clear (cloud-free) atmospheres, reaching almost $ 10\%$ \citep{StamHovenier2005, Sromovsky2005b}.

\subsection{Stellar Spectra}
\label{sec:stellar_types}

\begin{table}
\centering
\label{tab:stars}
\caption{Names and properties of stars used as input spectra \citep[from][]{Valdes2004}.}
\begin{tabular}{c c c c c}
\hline
 \mbox{Name} &  Spectral type & $T_{\textrm{eff}}$ (K)& log$_{10} g$ & [Fe/H] \\
 \hline
  HD 76644 & A7V & - & - & -  \\
  HD 33256 & F2V & 6442& 4.05 & -0.30 \\
  HD 115617 & G5V & 5590 & 4.23 & -0.03  \\ 
 HD 149661 & K2V & 5362& 4.56 & 0.01  \\
  HD 147379 & K7V & 3720 & 4.67 & 0.00 \\ 
   \hline
\end{tabular}
\end{table}

The structure and the intensity of Raman features strongly depend on the intensity of individual stellar lines and the overall shape of the stellar continuum. We use five different input spectra from the Indo-US Library of Coud\'{e} Feed Stellar Spectra\footnote{\url{https://www.noao.edu/cflib/}} \citep{Valdes2004}. These spectra cover the wavelength range between 3460~\AA\ and 9464~\AA, with the spectral resolution of $\sim 1$~\AA. We perform Gaussian smoothing and re-sampling to a uniform wavelength grid between 3500~\AA\ and 5500~\AA, with 2~\AA\ spacing. 

In order to properly model the redistribution of photons in wavelength-space due to Raman scattering, we need to start our radiative transfer calculation at shorter wavelengths than the wavelength range that we are interested in observing. For the strongest (ro-)vibrational Raman transitions of molecular hydrogen, Raman frequency shifts range between $\sim 4200 $ and $\sim 4800$~cm$^{-1}$ (pure rotational Raman transitions have much smaller shifts). Therefore, we need to start our radiative transfer at $\lambda \approx 3000$~\AA\ in order to properly treat the filling-in of lines at $\lambda \gtrsim 3500$~\AA. Since the \citet{Valdes2004} spectra do not cover wavelengths that far into the blue, we extrapolated their spectra with those from the Pickles atlas \citep{Pickles1998} in the wavelength range 3000-3500~\AA. We used the corresponding spectral type for each star, except in the case of A7V, which does not exist in the Pickles atlas, so we used the F0V Pickles spectrum instead. The Pickles spectra have a lower spectral resolution than the \citeauthor{Valdes2004} spectra, as they are reported on a wavelength grid with  $\Delta \lambda = 5$~\AA. This should not significantly affect the filling-in of lines longward of 3500~\AA, but it can cause the strength of Raman features in the low-resolution part of the spectrum to be underestimated. We report our main results---the geometric albedo spectra of planets around different spectral types of host stars---only for wavelengths $\lambda \geq 3500$~\AA, for which we have stellar spectra of higher resolution and for which the filling-in of lines by photons originating from shorter wavelengths is properly treated.   

The stellar spectra (normalized at 5000~\AA) are shown on the left-hand side of \autoref{fig:albedo}. The selected stars span the range of spectral types from A7V to K7V. \autoref{tab:stars} lists the names of the five stars whose spectra are used, along with their spectral type, effective temperature, surface gravity, and [Fe/H].

\section{Results and Discussion}
\label{sec:results}

\subsection{Geometric albedo spectra}

\begin{figure*}
\centering
\includegraphics[height=0.656\textheight]{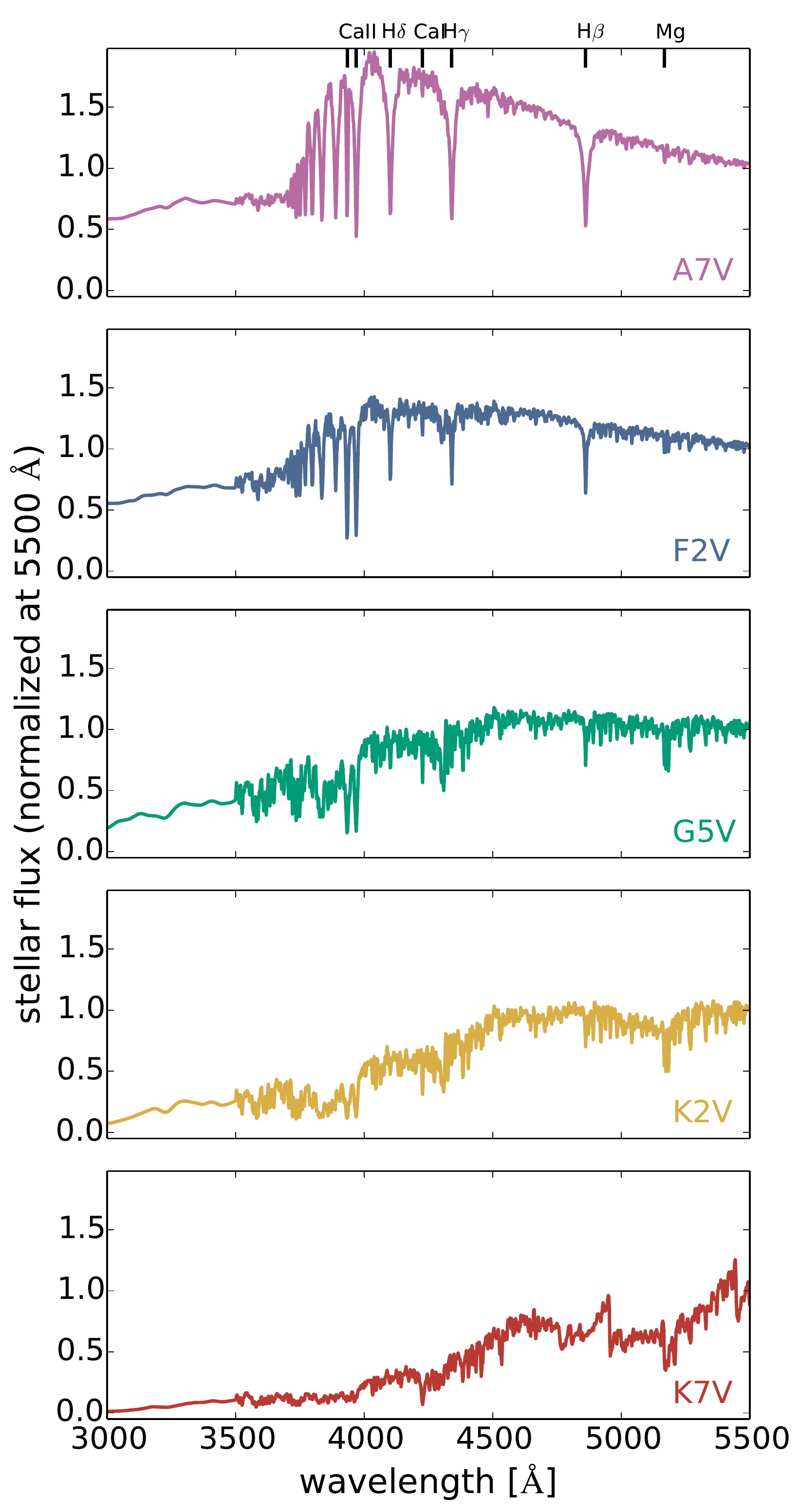}
\includegraphics[height=0.65\textheight]{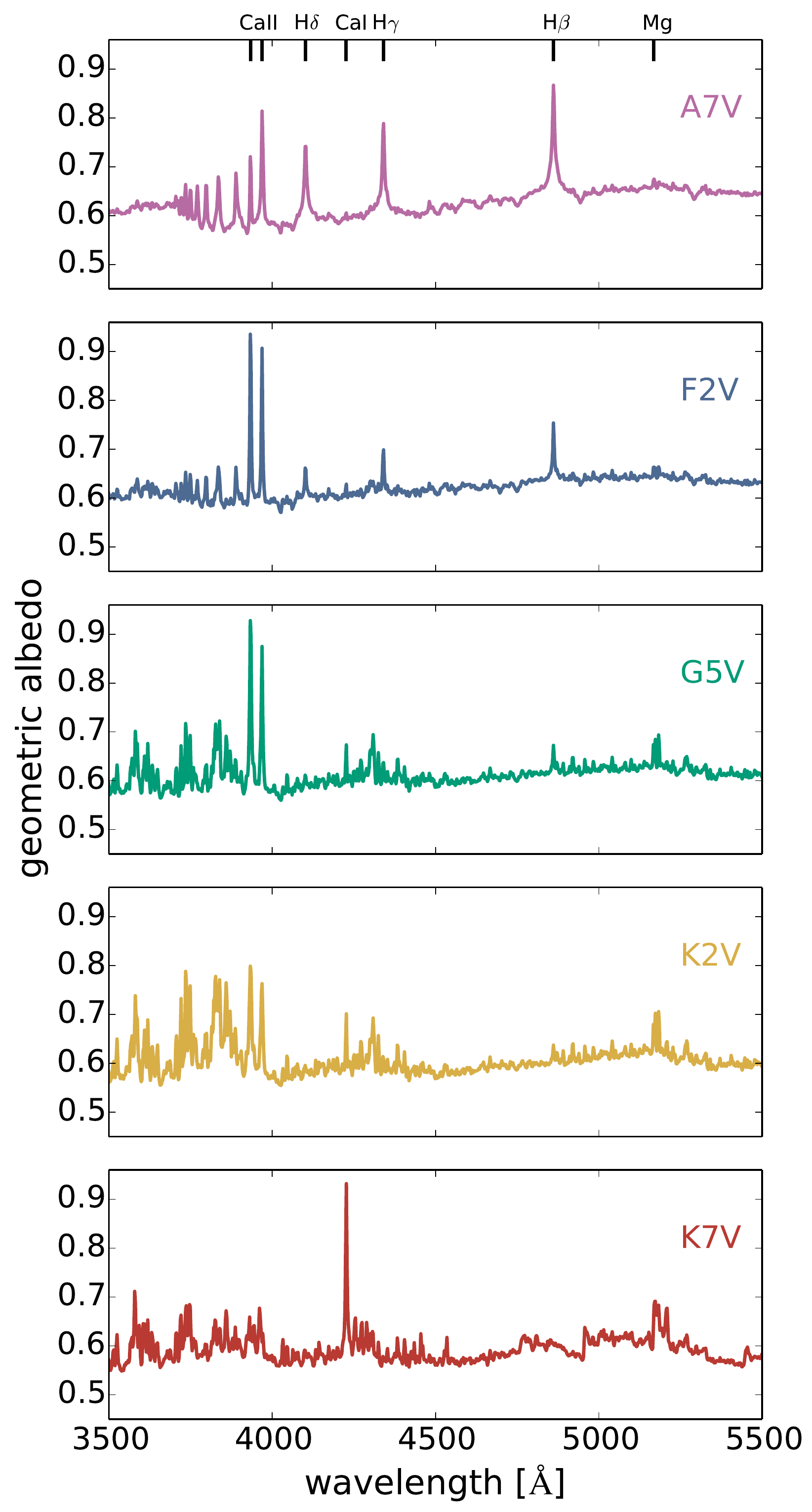}
\caption{Left: stellar spectra of five stars of different spectral types (see \autoref{tab:stars}). Bars on the top mark the positions of some of the most prominent spectral lines. Right: geometric albedo spectra calculated for a clear, 100~bar deep H$_2$/He atmosphere irradiated by the stellar spectra on the left-hand side. Due to the effects of Raman scattering, the albedo spectra contain peaks at wavelengths corresponding to prominent stellar lines.}
\label{fig:albedo}
\end{figure*}

\begin{figure*}
\centering
\includegraphics[height=0.75\textheight]{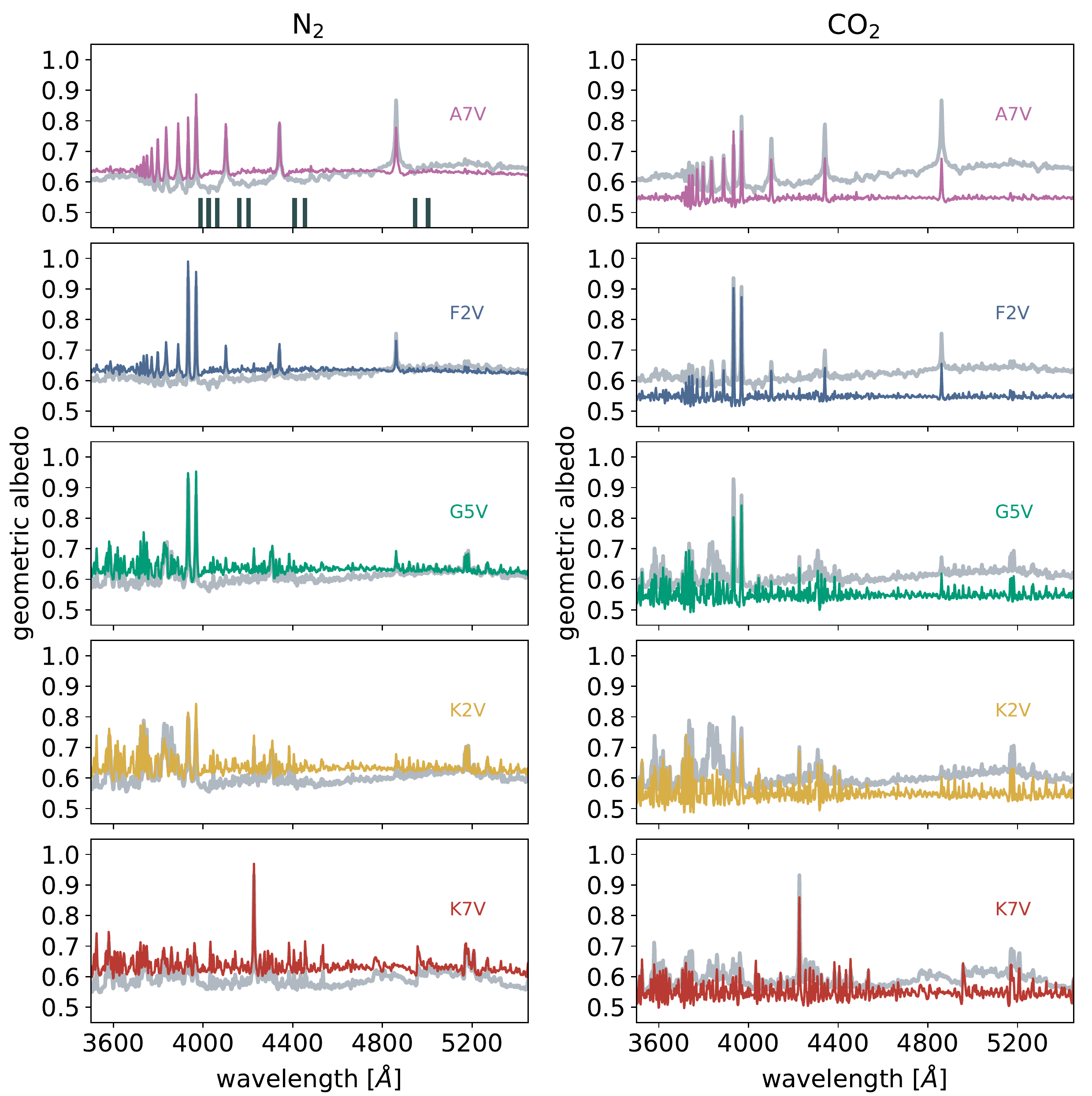}
\caption{Geometric albedo spectra calculated for clear, 100~bar deep atmospheres of pure N$_2$ (left-hand side) and CO$_2$ (right-hand side), irradiated by stellar spectra shown in \autoref{fig:albedo}. For comparison, the underlying gray lines show the corresponding H$_2$/He albedo spectra from \autoref{fig:albedo}. Dark bars in the top-left panel mark the central wavelengths of some of the most prominent hydrogen Raman ghost lines, corresponding to broad dips in the albedo spectra of hydrogen, which do not appear in the albedo spectra of heavy molecules at the same wavelengths.}
\label{fig:heavy_mols}
\end{figure*}

The geometric albedo spectra of the H$_2$/He model atmosphere, calculated for each of the five stellar spectra, are shown on the right-hand side of \autoref{fig:albedo}. Note that Rayleigh scattering alone produces a smooth albedo spectrum, and thus all the features seen in the albedo spectra are the result of Raman scattering. The atmospheric model is identical in all five cases shown in \autoref{fig:albedo} and all variations in the albedo spectra are due to differences in the stellar spectra. The overall shape of the stellar spectrum and the strength of individual stellar lines change from one spectral type to another, and they both affect the intensity of Raman features.

Albedo peaks are caused by photons that have been Raman-scattered into the wavelengths of absorption lines, mostly from shorter wavelengths. The presence of absorption lines in the stellar spectrum is therefore a necessary condition for Raman albedo peaks to arise. The most prominent albedo peaks appear at wavelengths corresponding to the strongest lines in the stellar spectrum, which vary across different spectral types. Raman features at short wavelengths are more pronounced in an albedo spectrum of an atmosphere irradiated by a hot star, whose blackbody flux peaks at shorter wavelengths, compared to an atmosphere around a cool star, which is much fainter at blue wavelengths. 

In the case of the earliest stellar type we consider, A7V, the most prominent features in the geometric albedo spectrum are associated with hydrogen Balmer lines: H$\beta$ at 4861~\AA, H$\gamma$ at 4340~\AA, and H$\delta$ at 4102~\AA. In F-type stars, the Balmer lines become less pronounced, and the H and K lines of ionized calcium, at 3968~\AA\ and 3933~\AA, respectively, become the dominant features. For G-type stars, the albedo spectrum is dominated by the Ca~II H and K lines\footnote{The peaks in the albedo spectrum for the G5V star, shown in \autoref{fig:albedo}, appear slightly less intense than the peaks produced by the solar spectrum in a similar planetary atmosphere shown in \citetalias{Oklopcic2016}. This is mainly the result of different spectral resolution---here we compute albedo spectra on a wavelength grid with spacing twice as large compared to what was used in \citetalias{Oklopcic2016} (2~\AA\ versus 1~\AA). Raman features are created by individual stellar lines, and thus their sharpness depends on the spectral resolution.}, and the Raman peaks corresponding to the Balmer lines become even fainter than in the case of F stars. The magnesium feature at $5167$~\AA\, and the Ca~I line at $4226$~\AA\ start to arise. In early K stars, the Ca~II lines are still quite strong and the features due to magnesium and calcium appear more prominently. They become the dominant features in albedo spectra calculated for late K-types.

In \autoref{fig:heavy_mols}, we show the geometric albedo spectra calculated for two additional atmospheric models, the atmospheres of pure N$_2$ (shown on the left-hand side) and pure CO$_2$ (on the right-hand side), irradiated by our five stellar spectra. The corresponding albedo spectra of the H$_2$/He atmosphere are shown for comparison in each panel in gray. To first order, the albedo spectra of N$_2$ and CO$_2$ atmospheres look very similar to those calculated for the H$_2$/He model, having a continuum value of $\sim$0.6 overlaid with Raman features at the wavelengths corresponding to prominent stellar lines. However, there are some notable differences arising from the fact that different molecules have different Rayleigh and Raman cross sections, as well as different values of Raman shifts.

Raman shift is the frequency offset imparted to Raman-scattered photons. It corresponds to the energy difference between the initial and the final state of the molecule. For pure rotational Raman transitions, the initial and the final state belong to the same vibrational level of the molecule ($\Delta v =0$), but differ in the rotational quantum number $J$ by two ($\Delta J = \pm 2$). For ro-vibrational transitions, the initial and the final vibrational levels of the molecule are different ($\Delta v \neq 0$), and the rotational quantum number can either change by two or remain the same ($\Delta J = 0,\pm 2$). The energy difference between different rotational states of a molecule is inversely proportional to its moment of inertia; hence, the energy states in heavy molecules are more closely spaced than in light molecules. For that reason, Raman frequency shifts are much smaller for heavy molecules, such as N$_2$ and CO$_2$, than for H$_2$. This explains why individual albedo peaks can have different intensity depending on the scattering molecule, even though they generally seem to be of comparable size in all three types of atmospheres. Raman ghosts, on the other hand, are much more pronounced in the H$_2$/He model (the central wavelengths of some of the most prominent ghost lines are indicated by gray bars in the top-left panel of \autoref{fig:heavy_mols}) than in the other two cases, in which they are difficult to discern by eye because the ghosts and albedo peaks are so close in wavelength that they almost overlap at this spectral resolution.

\subsection{The effects of atmospheric absorption}

The results shown in \autoref{fig:albedo} and \autoref{fig:heavy_mols} are computed for deep, clear atmospheres (i.e. atmospheres with large gas column density) of very simplified composition without any absorbers. These are the conditions in which the effects of Raman scattering are strongest, hence they produce the most optimistic predictions for the intensity of Raman features in the albedo spectra of exoplanets. 

Real atmospheres are expected to contain atomic and molecular absorbers, as well as clouds and hazes. The intensity of Raman features under these conditions is going to be reduced compared to our idealized case. In \citetalias{Oklopcic2016}, we explored how the strength of Raman albedo features changes with varying the radiation-penetration depth of the atmosphere (related to the effective altitude of opaque clouds), the reflective properties of the could deck, the surface gravity of the planet, and the temperature-pressure profile of the atmosphere. Here we expand on our previous work and investigate the effect of absorption by atomic species present in the atmosphere.

The main reason why the geometric albedo is expected to be relatively high at short wavelengths is that this wavelength range does not contain many absorption lines/bands of various atmospheric constituents; instead, scattering is the main source of opacity. At longer visible and infrared wavelengths, the atmospheric opacity is dominated by absorbers such as water or methane, and the expected geometric albedo is therefore small. 

However, there are a few notable absorption lines at short wavelengths associated with alkali metals sodium (Na) and potassium (K), which may affect the albedo spectrum of an exoplanet to a varying degree. The strength of the albedo absorption lines depends on the abundance of these atoms in monatomic form, as well as the ambient temperature and pressure. The mixing fractions of Na and K in giant planet atmospheres at high temperatures ($\gtrsim 1000$~K) are close to solar values, $\sim 10^{-6}$ for Na, and $\sim 10^{-7}$ for K \citep[see Figure 19 from][]{SharpBurrows2007}. Below $\sim 1000$~K the abundances start to rapidly decline due to the formation of condensates, which lock-in these elements and remove them from the gas phase. 

To demonstrate how the geometric albedo spectrum of an exoplanet might be affected by alkali absorption, we perform a new set of radiative transfer calculations using modified versions of the H$_2$/He atmospheric model, irradiated by an early K-type stellar spectrum. We include Na and K absorption as additional sources of opacity (along with Rayleigh scattering on H$_2$ and He, and Raman scattering on H$_2$) in our calculation of the single-scattering albedo at each wavelength, which is used as input for \textsc{disort} radiative transfer code. The calculation of absorption cross sections for Na and K is described in \autoref{app:alkali}. 

\begin{figure}
\centering
\includegraphics[width=0.45\textwidth]{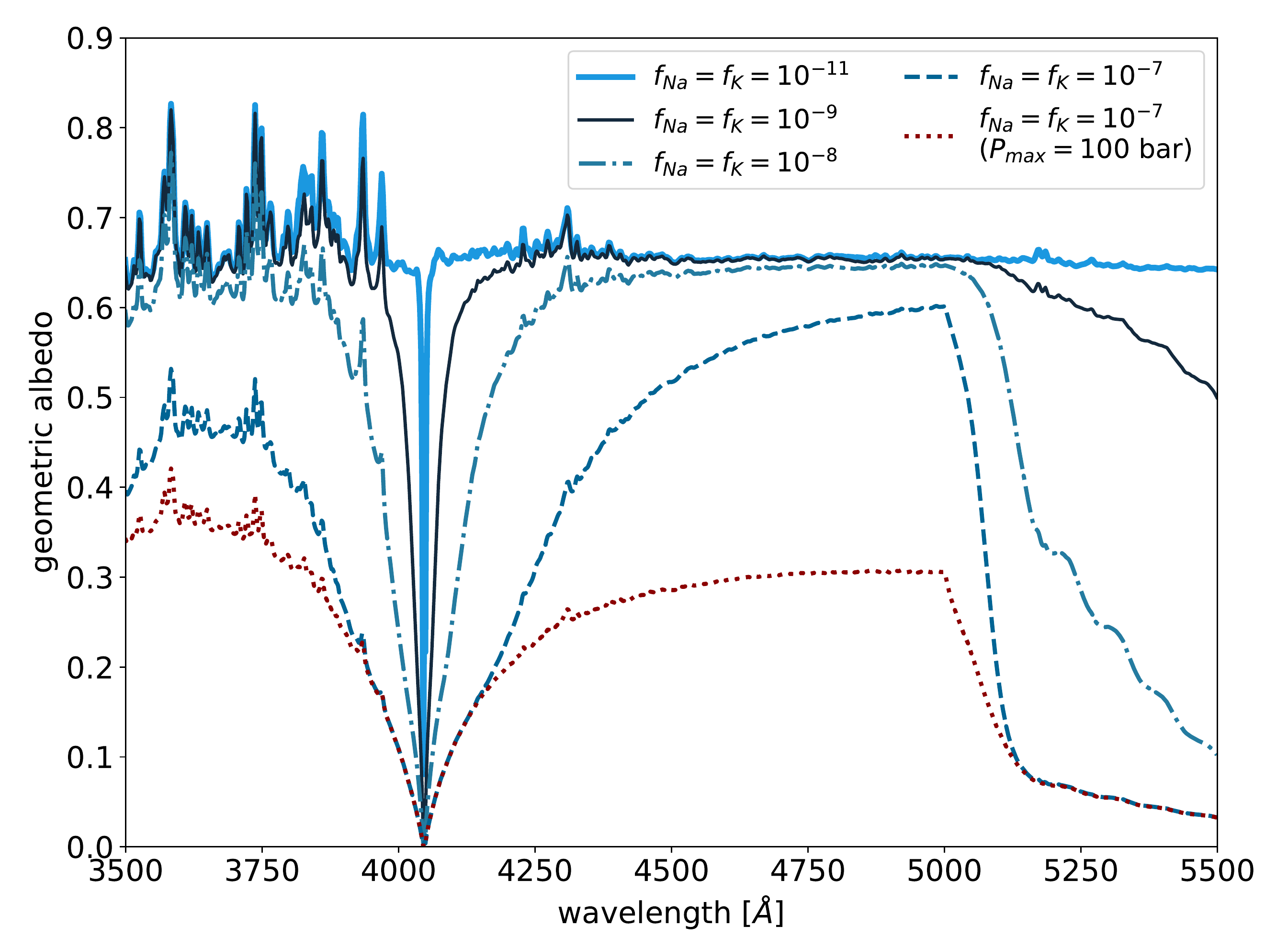}
\caption{Geometric albedo spectra of atmospheres containing different mixing fractions of Na and K atoms. The atmospheres are irradiated by the spectrum of an early K star.}
\label{fig:absorption}
\end{figure}

\autoref{fig:absorption} shows the albedo spectra of H$_2$/He-dominated atmospheres with sodium and potassium mixing ratios in the range $10^{-11}-10^{-7}$. The bottom boundary of the model atmosphere shown in red is at a depth of 100~bar. In atmospheres with lower fractions of alkali metals (presumably due to the formation of condensates), the bottom boundary is placed `by hand' at the pressure level of 1~bar and it is assumed to be a partially absorbing surface with a wavelength-independent\footnote{Aerosols are expected to have extinction curves that are relatively smooth functions of wavelength \citep[e.g.][]{deKok2011}, and hence should not produce spectral lines which might be confused with Raman features.} Lambertian albedo of 0.9. The bottom boundary properties are not calculated for any specific cloud model, they were selected to broadly represent two types of atmospheres---clear and cloudy---and to demonstrate that both the depth of the atmosphere and the mixing ratio of absorbers affect the strength of absorption lines which may preclude the formation of Raman spectral features.

The geometric albedos in \autoref{fig:absorption} show the effects of absorption lines of sodium at $3303$~\AA\ (which is not explicitly shown in the figure, but nevertheless it has an effect on the Raman-scattered light at longer wavelengths), of potassium at $4045$~\AA, and the Na-D resonance at $ 5890$~\AA, whose impact extends far away from the line center. For Na and K mixing fractions below $\sim 10^{-9}$, corresponding to atmospheric temperatures $\lesssim 800$~K \citep{SharpBurrows2007}, the absorption lines cause only a small change in the albedo spectrum compared to our idealized model without any absorption (\autoref{fig:albedo}). Therefore, alkali absorption can be safely ignored in moderately warm and cool atmospheres; however, it can inhibit the formation of prominent Raman spectral features in exoplanet atmospheres of very high temperature, such as those of hot Jupiters. 

As demonstrated in \citetalias{Oklopcic2016} (Figure 6), placing the cloud deck higher in the atmosphere suppresses the formation of Raman features more by effectively reducing the column of gas available for Raman scattering. If the atmosphere has substantial high-altitude hazes, radiation scattering and absorption on aerosols may mute Raman spectral features almost entirely. Our simplified cloud model is as just one example of how the presence of different types of condensates can shift the altitude of the planetary photosphere. In reality, aerosols can have a wide range of absorbing and scattering properties \citep[e.g.][]{Marley2013}. For example, it has recently been proposed that sulfur hazes, produced by photochemical destruction of H$_2$S, may act as important sources of absorption below $\sim 4000$~\AA\ in some giant exoplanets. This absorption can result in extremely small values ($\lesssim 0.1$) of the geometric albedo at short wavelengths \citep{Zahnle2016,Gao2017}; under such conditions the effects of Raman scattering would be practically unobservable.

\subsection{Observing the Raman features}

The geometric albedo spectra at short optical wavelengths could, in principle, be measured using a few different observational techniques: (1) secondary eclipse observations, similar to the \textit{STIS/HST} albedo measurements for HD~189733~b by \citet{Evans2013}; (2) high-resolution spectroscopy with large ground-based observatories, similar to the technique used by \citet{Martins2015}; and (3) direct spectroscopy using telescopes capable of achieving high contrasts at short optical wavelengths. There are no existing instruments in that category at the moment; however, a couple of concept studies for space-based missions equipped with coronagraphs and/or a starshades are under consideration, such as the \textit{Habitable Exoplanet Imaging Mission (HabEx)} and the \textit{Large UV/Optical/Infrared Surveyor (LUVOIR)}.

\begin{figure}
\centering
\includegraphics[width=0.5\textwidth]{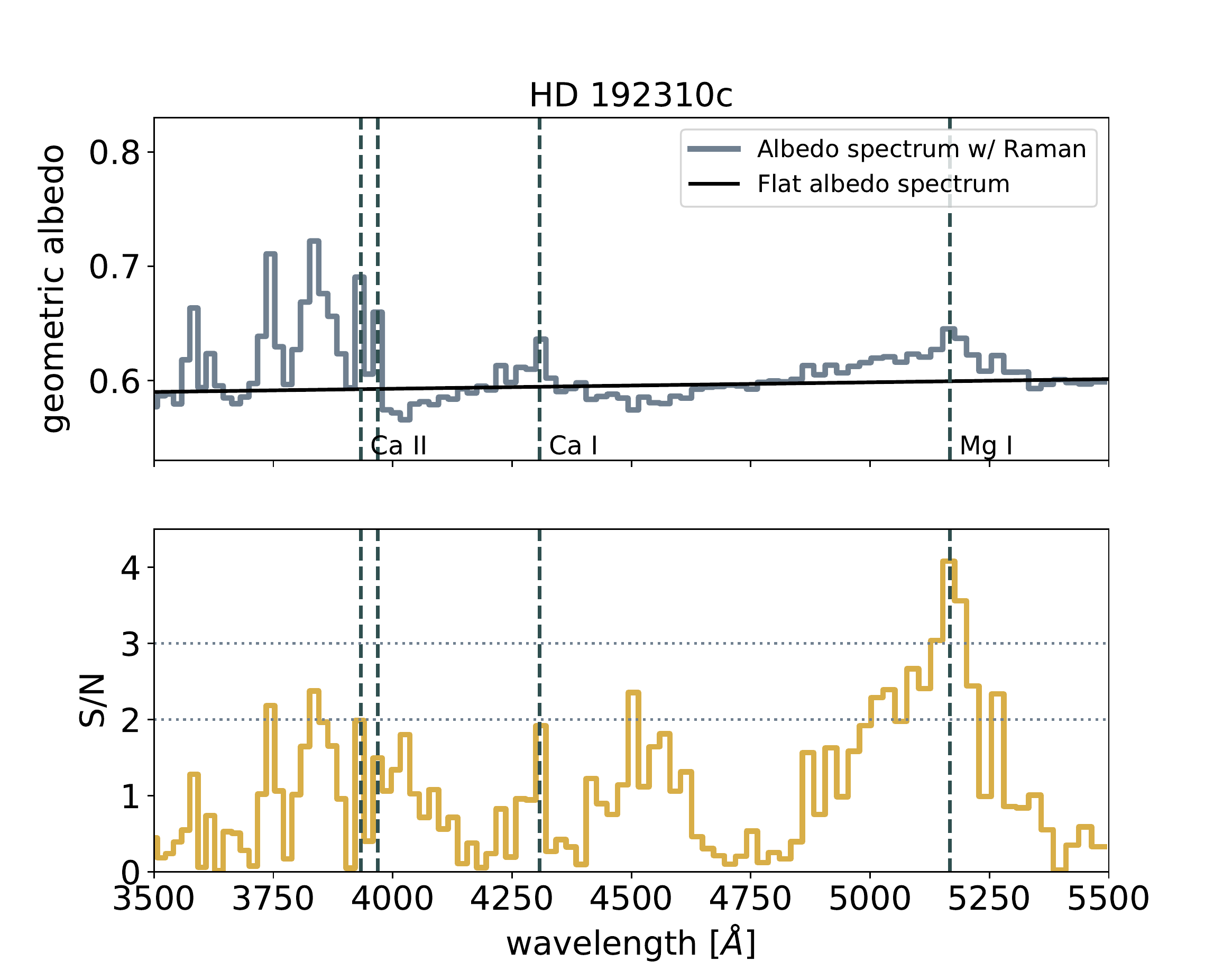}
\caption{Upper panel: the solid gray line shows the predicted geometric albedo spectrum for an H$_2$/He atmosphere on a planet with properties of HD 192310~c, observed at $R=200$ spectral resolution. The black line represents a smooth albedo spectrum without any Raman features. Observational signal is defined as the difference between these two curves. Lower panel: predicted S/N distribution as a function of wavelength for observations with a 5~m space telescope, for a total integration time of 100~hr, based on a modified version of the coronagraph noise model from \citet{Robinson2015}.}
\label{fig:snr_k2}
\end{figure}

 In \autoref{fig:snr_k2} we show the predicted signal-to-noise ratio (S/N) for direct detection of Raman features in the albedo spectrum of a nearby planet orbiting an early K star, HD~192310~c \citep{Pepe2011}. The properties of the planet-star system are listed in \autoref{tab:hd192310}. The low atmospheric temperature expected for this planet justifies the use of the albedo spectrum computed without any absorption from Na and K. S/N is calculated using the instrument noise model for a space-based coronagraph described by \citet{Robinson2015}, with some modifications: telescope diameter of 5 m (which is in the range of apertures proposed for \textit{HabEx}), spectral resolution of $R=200$, and a coronagraph capable of reaching the contrast level of 10$^{-10}$. We estimate S/N at which Raman features in the geometric albedo spectrum could be detected and distinguished from a smooth and featureless albedo spectrum of an atmosphere without any signatures of Raman scattering. We define `signal' as the difference between the photon count coming from a planet with an albedo spectrum with prominent Raman features (solid line in the upper panel of \autoref{fig:snr_k2}) and the photon count coming from a planet with a featureless albedo spectrum (dashed line). The predicted S/N distributions are computed for observations with the phase function value of $\phi=0.5$ (corresponding to the phase angle of $\alpha = 70^\circ$), and for total integration time of 100~hr. Individual Raman features can be distinguished from a flat albedo spectrum with S/N of 2-3. Observing the spectral signatures of Raman scattering in the atmospheres of nearby exoplanets, such as HD~192310~c, is going to be an ambitious, but feasible task for the telescopes of next generation.

\begin{table}
\centering
\caption{Properties of HD 192310 c (GJ 785 c) and its host star. References: (1) \citet{Pepe2011}, (2) \citet{Nayak2017}, (3) \citet{Fortney2007}, (4) \citet{Gray2006}, (5) \citet{Hog2000}, (6) \citet{Howard2012}.}
\begin{tabular}{c c c}
\hline 
Parameter & Value & Reference \\
\hline
  M$_\mathrm{pl}\sin{i}$ & 0.076 M$_\mathrm{Jup}$ & (1)\\
   R$_\mathrm{pl}$ & 0.75 R$_\mathrm{Jup}$& (2, 3)\\
  a & 1.18 au  & (1)\\ 
 T$_\mathrm{eff}$ & 185 K & (1) \\
 \hline
 spectral type & K2+V & (4) \\ 
 $V$ magnitude & 5.723 & (5) \\
  distance & 8.9 pc & (6) \\
   \hline
\end{tabular}
\label{tab:hd192310}
\end{table}

\begin{figure}
\centering
\includegraphics[width=0.48\textwidth]{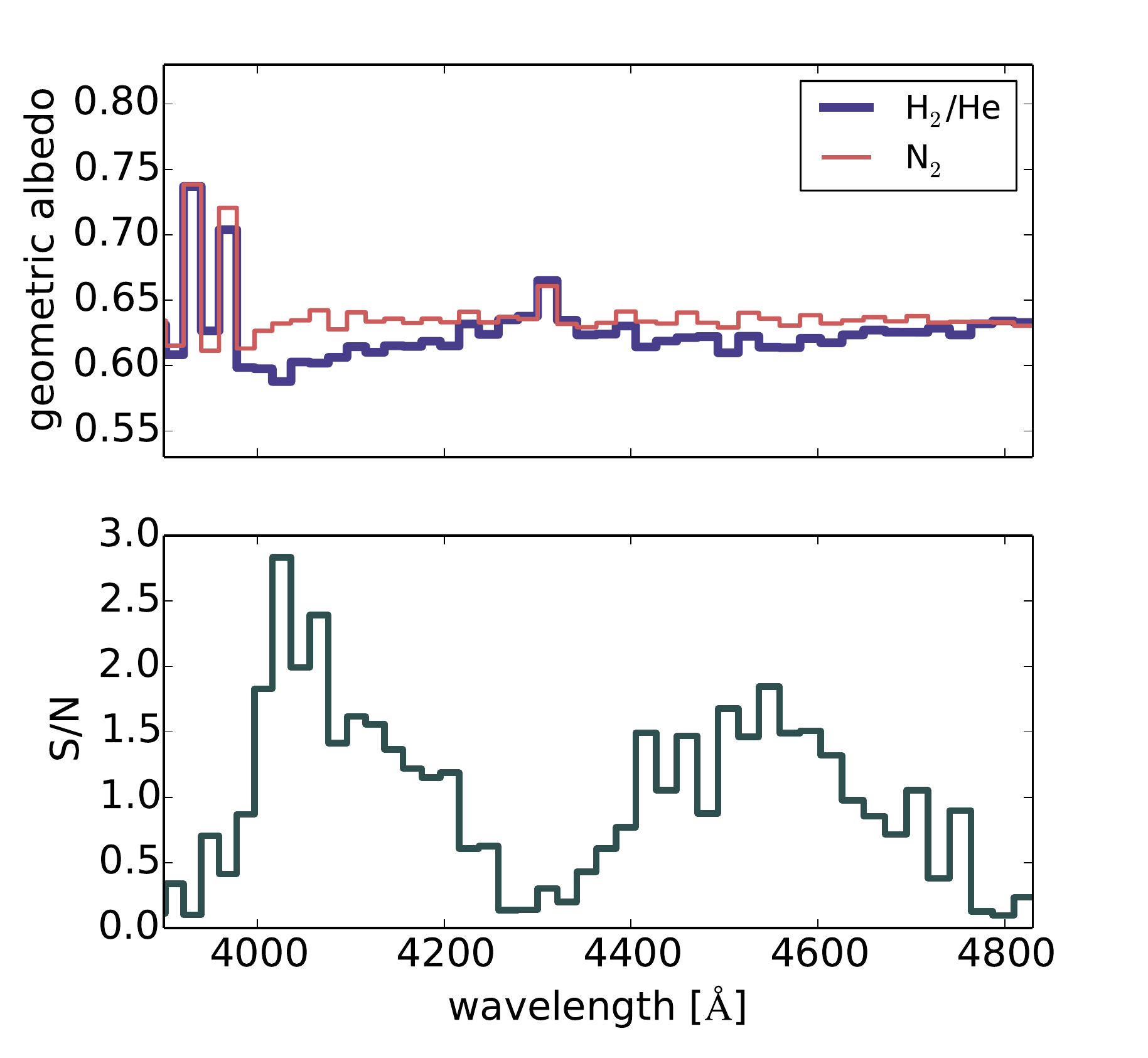}
\caption{Upper panel: geometric albedo spectra for H$_2$/He and N$_2$ atmospheres, irradiated by G5V stellar spectrum, observed at $R=200$ spectral resolution. These spectra are computed for deep, clear atmospheres in which Raman features are strongest. Lower panel: S/N distribution for distinguishing between H$_2$ and N$_2$ atmospheres on an Earth-sized planet orbiting at 1~au separation from the host star, at a distance of 7~pc. We assume a telescope diameter of 10~m and the total integration time of 400~hr.}
\label{fig:snr_earth}
\end{figure}

Even more ambitious goal would be detecting the Raman features in Earth-sized planets and using them to distinguish between atmospheres dominated by heavy or light molecules. In \autoref{fig:snr_earth}, we show the albedo spectra of an H$_2$/He atmosphere and a pure N$_2$ atmosphere, both irradiated by the G5V spectrum (as in Figures \ref{fig:albedo} and \ref{fig:heavy_mols}). Using the noise model described above, we estimate the S/N to which these two atmospheres can be distinguished (i.e. signal is defined as the difference between photon counts computed for the two types of atmosphere) on an Earth-sized planet, at 1~au separation from a Sun-like star located at a distance of 7~pc. The lower panel in \autoref{fig:snr_earth} shows the S/N distribution for 400-hr observations with a $D=10$~m telescope (which is in the range of sizes proposed for \textit{LUVOIR}) at spectral resolution of $R=200$. Two broad `bumps' in the S/N distribution arise from two sets of H$_2$ Raman ghost lines of the Ca~II H and K lines: ghosts originating from rotational Raman transitions (the closer bump, ranging from ~4000 to ~4200 \AA) and ghosts produced by ro-vibrational transitions (the bump between ~4400 and ~4800 \AA). The strongest H$_2$ ghost feature appears at ~4030 \AA\, due to two overlapping rotational Raman ghosts of H and K lines.

The prospects for observing Raman features in the albedo spectrum of a planet depend on both the strength of the features and on the brightness of the host star and the planet in the part of the spectrum where the features appear. \autoref{fig:snr_k2} also illustrates that in some case, especially in late-type stars, moderate-size Raman features (such as the Mg~I albedo peak) can result in greater S/N than the most prominent albedo peaks, if the latter appear in a wavelength range in which the star is intrinsically faint. For G stars and earlier stellar types, the regions of highest S/N generally coincide with the most prominent Raman peaks.

\section{Conclusions}
\label{sec:conclusions}

Observations of the reflected light and the albedo spectra of exoplanets at short optical wavelengths ($3000 \lesssim \lambda \lesssim 6000$~\AA) can provide insight into atmospheric properties. Contrary to a common assumption, the albedo at these wavelengths is not solely determined by Rayleigh scattering. Observations at moderate spectral resolution (R$=10^2-10^3$) can detect the spectral signatures of Raman scattering, which carry information about the depth of the atmosphere, its temperature, and composition.

Raman features do not only depend on the properties of the planetary atmosphere---they also depend on the shape of the stellar spectrum irradiating the planet. In this paper, we explore how the positions (i.e. wavelengths) and the intensities of different Raman features change due to the differences in the incident stellar light across different spectral types of host stars. For each stellar type, ranging from late A  to late K stars, we identify the most promising Raman features that could be observed with the next generation of telescopes.  

The recommended strategy for detecting and measuring Raman features in the albedo spectra of exoplanets would be to obtain moderate-resolution ($R=10^2 -10^3$) spectra over the wavelength range containing strong stellar lines, such as the Balmer lines or the Ca~II H and K lines. The most suitable stellar types are the early types, which show strong Raman peaks and are also bright at short wavelengths. Late stellar types (such as our K7V case or later), although more numerous at close distance from the Sun, are intrinsically very faint at the wavelengths at which the most notable Raman peaks appear, thus making the observations of Raman features very challenging.

\acknowledgments
AO acknowledges support from an Institute for Theory and Computation Fellowship.  AO and CMH were supported by NASA, the U.S. Department of Energy, the David \& Lucile Packard Foundation, and the Simons Foundation. KH acknowledges support from the Swiss National Science Foundation, the PlanetS NCCR and the MERAC Foundation.

\software{\textsc{disort} \citep{Stamnes1988}}

\bibliographystyle{yahapj}
\bibliography{refs_raman}

\appendix
\section{Rayleigh and Raman Scattering Cross Sections for CO$_2$}
\label{app:co2}

 In order to compute CO$_2$ cross sections for Rayleigh scattering and the rotational Raman scattering, we use the approximate formulae from \citet{Dalgarno1962}. Because CO$_2$ has a relatively complicated structure of vibrational transitions, we consider only rotational Raman transitions in this work and assume all molecules are in the ground electronic and vibrational state. 
 
For a transition between the initial molecular state of CO$_2$, characterized by the rotational quantum number $J_i$, and the final state characterized by $J_f$, the cross section is given by the following formulae.
\begin{itemize}
\item For Rayleigh scattering ($J_i=J_f$):
\begin{equation}
Q(J_i;J_f) = \frac{128\pi^5}{9\lambda^4}\left[3\alpha^2 +\frac{2}{3}\frac{J_i(J_i+1)}{(2J_i-1)(2J_i+3)} \gamma^2\right] \ \mbox{.}
\end{equation}
\item For rotational S-branch transitions ($J_f=J_i+2$):
\begin{equation}
Q(J_i; J_f) = \frac{128\pi^5}{9\lambda^{\prime}\lambda^3}\frac{(J_i+1)(J_i+2)}{(2J_i+3)(2J_i+1)}\gamma^2 \ \mbox{.}
\end{equation}
\item For rotational O-branch transitions ($J_f=J_i-2$):
\begin{equation}
Q(J_i; J_f) = \frac{128\pi^5}{9\lambda^{\prime}\lambda^3}\frac{J_i(J_i-1)}{(2J_i-1)(2J_i+1)}\gamma^2\ \mbox{,}
\end{equation}
\end{itemize}
where $\lambda$ and $\lambda^\prime$ are the outgoing and the incident wavelength, respectively. The incident and outgoing wavelengths are related by $\lambda^{\prime -1} = \lambda^{-1} + \Delta \nu$, where $\Delta \nu$ is the Raman shift of the transition in question, which is related to the energy difference between the initial and final state of the molecule involved in the scattering process. $\alpha$ and $\gamma$ are the polarizability and the polarizability anisotropy of CO$_2$ at the equilibrium separation of nuclei, which have values of $\alpha = 2.3353 \times 10^{-24}$ cm$^{3}$ and $\gamma = 1.7915 \times 10^{-24}$ cm$^3$ \citep{MorrisonHay1979}. The differences between rotational energy levels are computed using the values of rotational constants for CO$_2$, $B=0.3902$~cm$^{-1}$ and $D=0.12\times 10^{-6}$~cm$^{-1}$, from \citet{Herzberg1953}.

\section{Absorption cross sections for sodium and potassium}
\label{app:alkali}

The absorption cross section for an atomic transition from a lower state $l$ to an upper state $u$ is given by
\begin{equation}
\sigma_{lu} (\nu) = \frac{\pi e^2}{m_e c}f_{lu}\phi_{\nu} \ \mbox{,}
\end{equation}
where $e$ is the elementary electric charge, $m_e$ is the electron mass, $c$ is the speed of light, $f_{lu}$ is the oscillator strength of the transition, and $\phi_\nu$ is the line profile, normalized so that $\int \phi_\nu d\nu=1$.

We adopt the Voigt line profile, which is a convolution of a Gaussian and a Lorentzian profile, computed using the \textsc{python} code described in \citet{Hill2016}\footnote{\url{http://scipython.com/book/chapter-8-scipy/examples/the-voigt-profile/}}. The line profile is calculated using the real part of the Faddeeva function $\omega(z)$ as
\begin{equation}
V(\nu-\nu_0; \sigma, \gamma)=\frac{\operatorname{Re}[\omega(z)]}{\sigma \sqrt{2\pi}} \ \mbox{,} 
\end{equation}
where $z=\frac{\nu-\nu_0+i\gamma}{\sigma\sqrt(2)}$, $\nu_0$ is the line center frequency, $\sigma$ is the standard deviation of the Gaussian profile, and $\gamma$ is the half-width at half-maximum (HWHM) of the Lorentzian.

The Gaussian part of the line profile originates from thermal Doppler broadening. Its HWHM is given by
\begin{equation}
\alpha= \nu_0\sqrt{\frac{2\ln{(2)}k_BT}{mc^2}} \ \mbox{.}
\end{equation}
It is related to the Gaussian standard deviation $\sigma$ through $\alpha = \sigma \sqrt{2\ln{2}}$.

The dominant part of the Lorentzain line profile is caused by pressure broadening. We follow the approach described in \citet{Schweitzer1996} and model the pressure broadening of atomic lines in terms of the van der Waals (vdW) interaction between neutral, unpolarized particles. This interaction  leads to a Lorentzian line profile with a HWHM of
\begin{equation}
\gamma = \frac{17}{2}C_6^{2/5}v_{\mathrm{rel}}^{3/5} n_{\mathrm{per}} \ \mbox{,}
\end{equation}
where $C_6$ is the vdW interaction constant, $v_{\mathrm{rel}}$ is the relative velocity between the absorbing atom (Na or K) and the perturber, and $n_\mathrm{per}$ is the number density of the perturber. For $v_{\mathrm{rel}}$, we use the expression for the mean relative speed between particles described by the Maxwell-Boltzmann distribution
\begin{equation}
\langle v_{\mathrm{rel}} \rangle = \sqrt{\frac{8k_BT}{\pi \mu}}\ \mbox{,}
\end{equation}
where $\mu = m_1m_2/(m_1 + m_2)$ is the reduced mass of the two particles.

The interaction constant is given by
\begin{equation}
C_6 = 1.01\times 10^{-32}(Z+1)^2\frac{\alpha_p}{\alpha_H}\left[\frac{E_H^2}{(E-E_l)^2} - \frac{E_H^2}{(E-E_u)^2} \right] \mbox{cm$^6$s$^{-1}$} \ \mbox{,}
\end{equation}
where $Z$ is the electric charge of the absorber, $\alpha_p$ and $\alpha_H$ are the polarizabilities of the perturber and the hydrogen atom, respectively. Since our model atmosphere is mostly made of H$_2$, we assume that this molecule is the dominant perturber and that all pressure broadening is due to interactions of the absorbing species with H$_2$. The polarizabilities relevant for our study are:
\begin{eqnarray}
\alpha_H &=& 0.666793 \times 10^{-24} \ \mbox{cm$^3$}  \mbox{,}\\
\alpha_{H_2} &=& 0.806 \times 10^{-24} \ \mbox{cm$^3$} \mbox{.} 
\end{eqnarray}
$E$ and $E_\mathrm{H}$ denote the ionization potential of the absorbing species and of the hydrogen atom, respectively:
\begin{eqnarray}
E_\mathrm{H} &=& 13.6 \ \mbox{eV}  \mbox{,}\\
E_\mathrm{Na} &=& 5.139 \ \mbox{eV}  \mbox{,}\\
E_\mathrm{K} &=& 4.341 \ \mbox{eV}  \mbox{.}
\end{eqnarray}
$E_l$ and $E_u$ are the excitation energies of the lower and upper level involved in the transition. There are four strong transitions (two doublets) in the wavelength range considered in our analysis, listed in \autoref{tab:lines}. All atomic parameters are taken from the \textit{NIST} Atomic Spectra Database\footnote{\url{http://physics.nist.gov/PhysRefData/ASD/lines_form.html}}. 

\begin{table}
\centering
\label{tab:lines}
\caption{Spectral line parameters.}
\begin{tabular}{l c l c l}
\hline
 Atom &  Wavelength [nm] & f$_{ik}$ & E$_i$ [eV] & E$_k$ [eV] \\
 \hline
  Na I & 330.3320 & 0.009 & 0.0 & 3.753322 \\
  Na I & 330.3930 & 0.00446 & 0.0 & 3.752628 \\
  K I & 404.5279 & 0.00569 & 0.0 & 3.0649066 \\
  K I & 404.8351 & 0.00263 & 0.0 & 3.062581\\
   \hline
\end{tabular}
\end{table}

Strong resonances of sodium and potassium at 5890 and 7700~\AA\ can be important sources of opacity in exoplanet atmospheres at wavelengths far away from the line centers. Even though we do not extend our analyzed wavelength range beyond 5500~\AA, we include the absorption wings of the sodium-D resonance, which can affect the albedo spectra at wavelengths $\gtrsim 5000$~\AA. Describing the wings of these resonance lines with Lorentzian profiles has been shown to be inadequate. The Lorentz theory relies on the validity of the impact approximation (i.e. assuming instantaneous collisions between atoms and perturbers), which is not satisfied far from the line center of strong resonance lines. More rigorous treatments of the atom-perturber interaction produce line profiles with sharper cutoffs, stronger blue wings with satellite features, and a pronounced red/blue asymmetry \citep{Burrows2003,Allard2012}. 
We use the Na-D line profile from \citet{Allard2012}, calculated for an atmospheric density of $n (H_2)= 10^{19}$~cm$^{-3}$ and temperature 1500~K, made available by the authors in tabular form. Since the line profile is provided in terms of an absorption coefficient in arbitrary units, we re-scaled it so that the values in the blue wing match the cross section profile (in cm$^2$) from \citet[Figure 6]{Burrows2003}. The validity of the impact approximation (and consequently, the use of the Lorentzian profile in the line wings) for weaker sodium and potassium lines, such as those listed in \autoref{tab:lines}, can also be called into question. This could be relevant in atmospheres of high mixing ratios of Na and K, which produce relatively broad absorption features. Calculating more accurate cross section profiles is beyond the scope of this paper; therefore we use the vdW Lorentzian profiles, but note that this may not be completely justified.


\end{document}